\documentclass[12pt]{article}
\usepackage{times}
\usepackage{geometry}
\usepackage[utf8]{inputenc}
\usepackage{enumitem,amssymb}
\usepackage{ragged2e}
\usepackage{natbib}
\usepackage{graphicx}
\newlist{thematic}{itemize}{8}
\setlist[thematic]{label=$\square$}
\usepackage{pifont}

\usepackage{amsmath}    
\usepackage{amssymb}    

\begin{document}
\raggedright
\huge
Astro2020 Science White Paper \linebreak

Dynamical Processes in the Planet-Forming Environment
\linebreak
\normalsize


\noindent \textbf{Thematic Areas:} \hspace*{60pt} $\square$ Planetary Systems \hspace*{10pt} 
 $\boxtimes$ Star and Planet Formation \hspace*{20pt}\linebreak
$\square$ Formation and Evolution of Compact Objects \hspace*{31pt} $\square$ Cosmology and Fundamental Physics \linebreak
  $\square$  Stars and Stellar Evolution \hspace*{1pt} $\square$ Resolved Stellar Populations and their Environments \hspace*{40pt} \linebreak
$\square$    Galaxy Evolution   \hspace*{45pt} $\square$             Multi-Messenger Astronomy and Astrophysics \hspace*{65pt} \linebreak

\textbf{Principal Author:}

Name: Peregrine M. McGehee
 \linebreak
Institution: College of the Canyons
 \linebreak
Email: peregrine.mcgehee@gmail.com
 \linebreak
Phone: +1 (626) 993-4199
 \linebreak

\textbf{Co-authors:} \\
Alan Boss (Carnegie Institute of Washington),
Laird Close (University of Arizona),
Sarah Dodson-Robinson (University of Delaware),
Dimitri Mawet (California Institute of Technology),
Andrew Szentgyorgyi (Harvard-Smithsonian Center for Astrophysics)
  \linebreak

\textbf{Abstract:}
The transfer of circumstellar disk mass and momentum onto the protostar and out into the
environment occurs via a variety of mechanisms including magnetospheric accretion,
jets, outflows, and disk winds.  The interplay of these processes determine both the 
conditions under which planet formation occurs and the lifetime of the disk.
Metallic emission lines, along with the Balmer series of hydrogen, probe the
kinematics of gas within the planet-forming and central regions of circumstellar disks.
High-spectral resolution study of these emission lines provides critical information on
mass and momentum loss, turbulence, and disk wind origins. 

\pagebreak

\section{Introduction} 

The formation process of planets is constrained by the kinematics within the
inner circumstellar disk. 
Indeed, 
current models of protoplanetary disk evolution suggest that the lifetime of the disk in the planet-forming region 
is mainly defined by viscous evolution
(accretion of gas onto the central star) and photoevaporation driven winds
(heating of disk gas to thermal escape velocities by the central star)
\citep{Pascucci2011}. 
In this science white paper we describe how sensitive and high spectral resolution 
studies of Balmer and metallic emission lines reveal critical insights into the planet forming environment
including mass and momentum loss, turbulence, and disk wind origins.  In particular we
examine the significant information found by detailed observations of optical forbidden lines. Additional optical spectral features arising in the star--disk environment
include Balmer and metallic emission lines arising from magnetospheric accretion
columns and forbidden emission lines generated in the shocked regions associated with jets and outflows.

\begin{figure}[ht]
    \centering
    \includegraphics[width=0.6\textwidth]{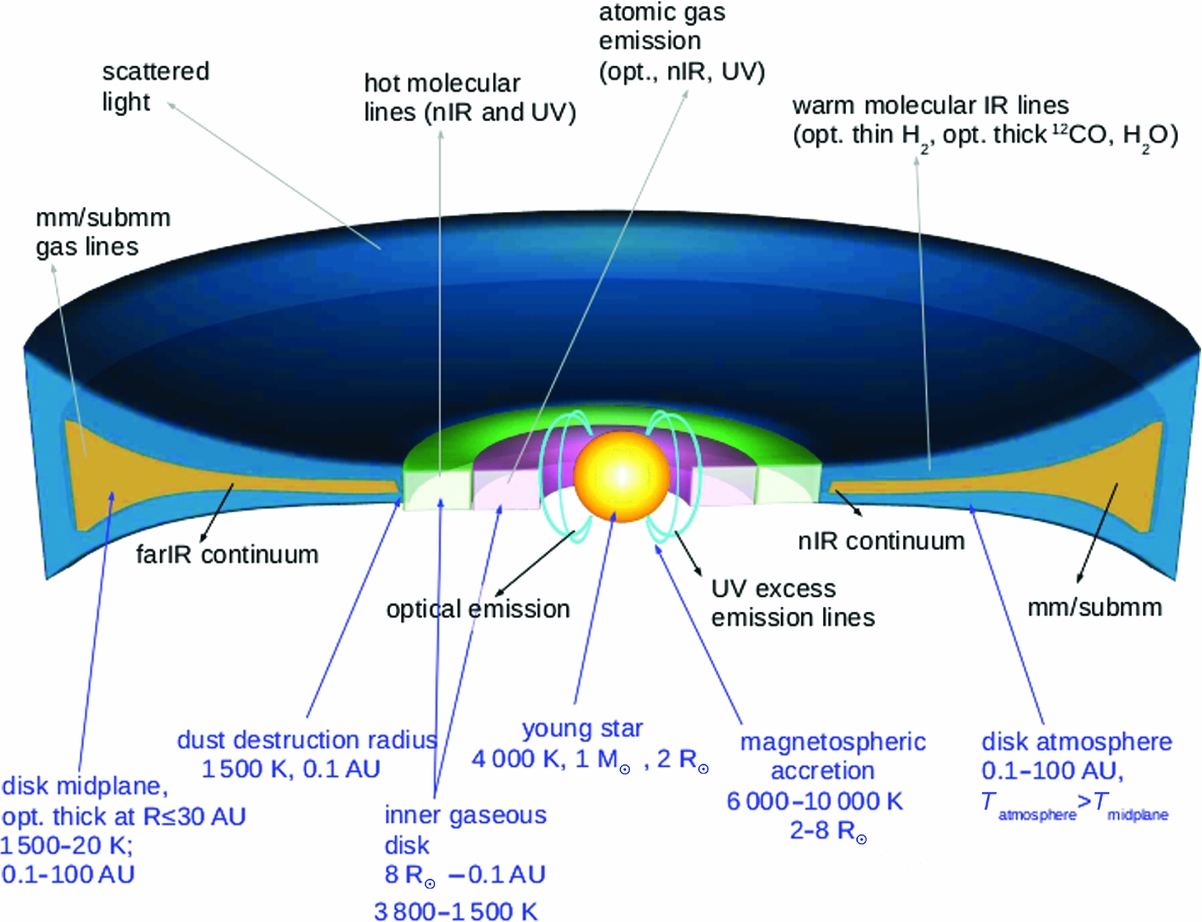}
    \caption{\label{fig:substructure}
From \citet{Sicilia-Aguilar2016}. 
A cartoon of the observations and the parts of the disk that they trace, taking as example a young solar analogue. Although
observations trace very different regions and processes in the disk, we need to keep in mind that they are all connected through the disk
itself. Note that the complexity of the disk is highly reduced for clarity (for instance, not all the tracers become optically thick at the same
location/depth). Not to scale.
    }
\end{figure}
In their in-depth consideration of multi-wavelength studies of circumstellar disks, see Figure \ref{fig:substructure}, 
\citet{Sicilia-Aguilar2016} point out that optical spectroscopy is well suited to target the physics
of accretion, photoevaporation, outbursts, activity, and binarity. Within the set of {\it Burning Questions}
listed by \citet{Haworth2016} in the context of {\bf Grand Challenges in Protoplanetary Disc Modelling}, those addressable by optical studies include:
\begin{itemize}
\item What are the main drivers of global disc evolution? In particular, what is the main driver of the mass accretion rate in protoplanetary discs?
\item Alongside magnetic fields, what other processes govern or control the launching of jets and outflows?
\item What is the effect of environment on protoplanetary disc evolution? For example, discs close to O stars are clearly heavily disrupted by high energy photons (we observe such systems as proplyds), but what is the role of comparatively modest radiation fields?
\item What are the possible initial conditions of class I/II/III discs and how do they influence the subsequent evolution? In particular, how does the early evolution of discs affect the chemistry and grain distribution? What is inherited from the star-formation process?
\item How turbulent are protoplanetary discs?
\item What is the process by which a protoplanetary disc becomes a debris disc? Transition discs; those with inner holes, are typically attributed to the action of photoevaporation by the host star, or planets. But which, if either, of these is the dominant process? Are there other processes that contribute significantly to disc dispersal, such as magneto-thermal winds? What are the initial conditions of debris disc models?
\end{itemize}

The profiles of forbidden emission lines arising from circumstellar disks
can be decomposed into separate low and high velocity 
components, of which the former can be interpreted following \citet{Simon2016} as a combination of a narrow and broad Gaussian features 
(Figure \ref{fig:example_lines}). 
The high-velocity component (HVC) is typically blueshifted by 
50--200 km s$^{-1}$
 with respect to the stellar velocity and {\bf traces collimated jets} that have been spatially resolved at distances from 50 to several 100 AU from the star.
The LVC is formed from a broad, centrally peaked component attributed to 
{\bf gas arising in a warm disk surface in Keplerian rotation}  (with FWHM between 40 and 60 km s$^{-1}$) and a narrow component (with FWHM  $\sim$10 km s$^{-1}$ and small blueshifts of $\sim$2 km s$^{-1}$).
The narrow component {\bf arises in a cool ($<$ 1000 K) molecular wind} as the peak velocities of various forbidden lines are inversely proportional to their respective critical densities – as expected where the flow accelerates as it rises from the disk.

\section{Main Science Themes}
 
The phenomena associated with the inner disk and circumstellar environment impact the planet formation process in differing ways and have associated observational tracers accessible by optical spectroscopy.  In this context we discuss magnetospheric accretion, disk winds, jets and outflows, and the shocked interaction regions between jets and the ISM.

\subsection{Magnetospheric Accretion}

\begin{figure}[ht]
    \centering
    \includegraphics[width=0.4\textwidth]{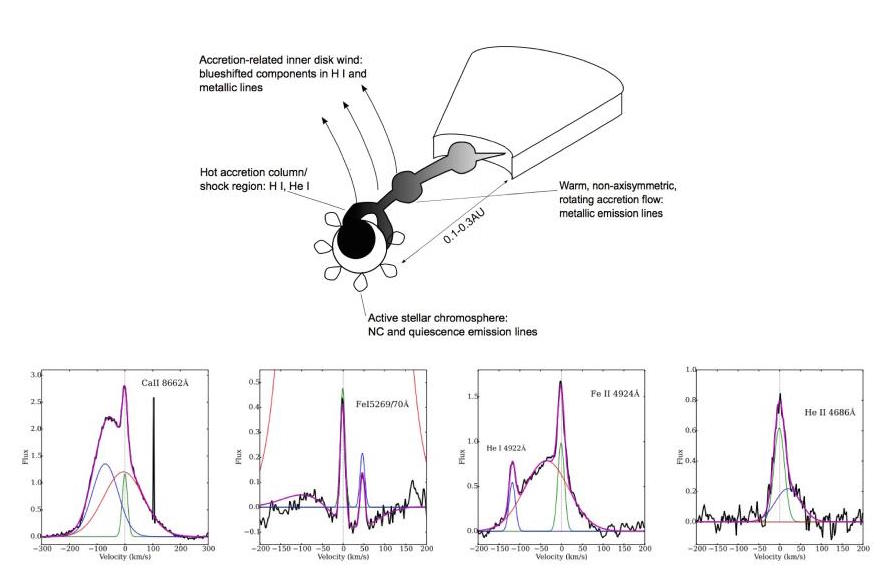}
    \includegraphics[width=0.4\textwidth]{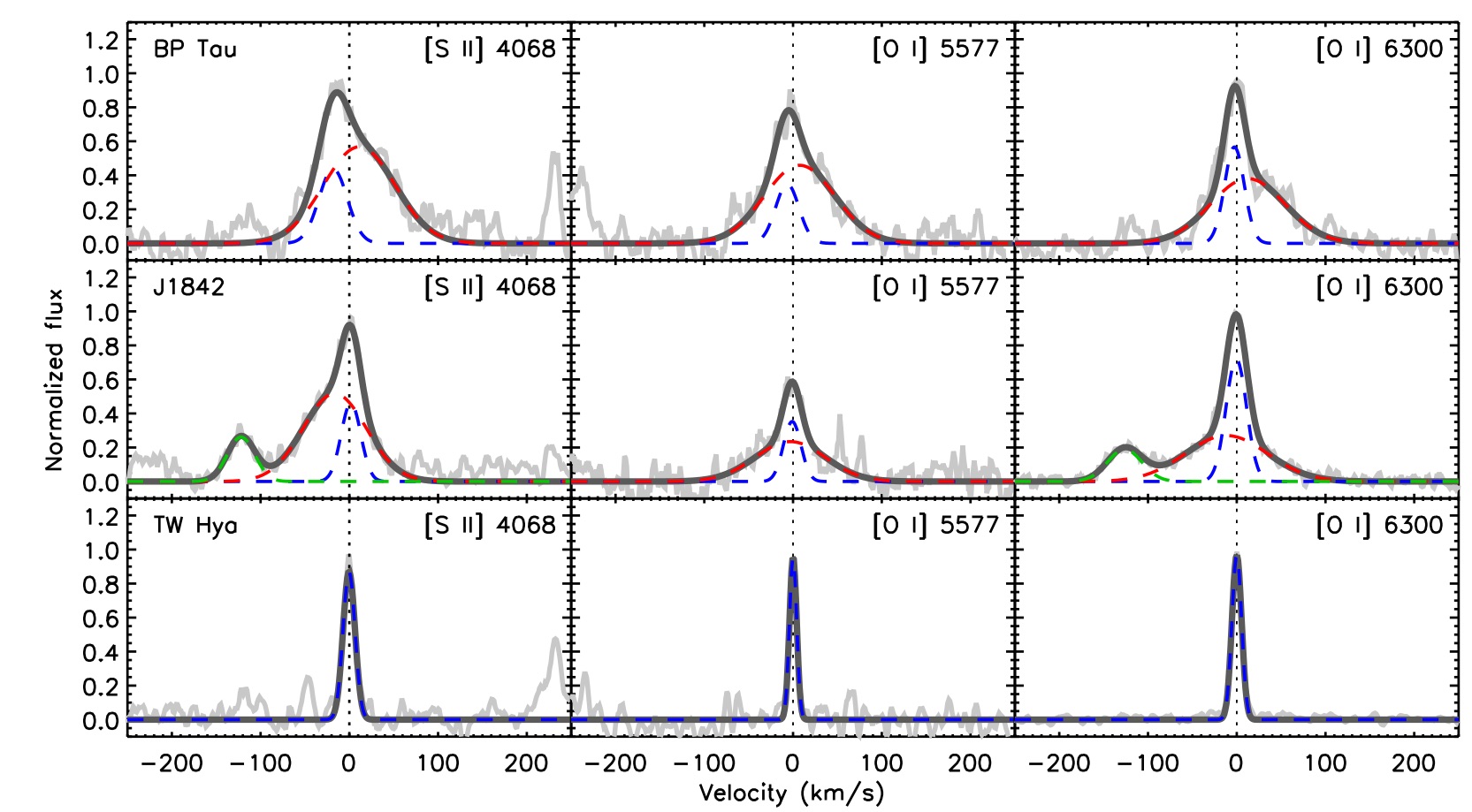}
    \caption{\label{fig:example_lines}
Examples of emission line observations probing accretion ({\it left}, from \citet{Sicilia-Aguilar2015}),
and disk winds ({\it right}, from \citet{Fang2018}) physics, with the latter utilizing metallic forbidden lines.
Lines associated with the magentospheric accretion column are shown here with decomposition
into Gaussian components. The optical forbidden lines probing the disk wind and outflow launching environments, such as  [S II] 4068{\AA}, [O I] 6300{\AA}, and [O I] 5577{\AA} show distinct high and low velocity features.}
\end{figure}

The process of magnetospheric accretion transfers mass and momentum from the disk onto the star and directly affects
the inner regions of the cirumstellar disk.
The {\bf central ($r < 0.1$ AU) regions} containing magnetospheric accretion columns and the launching sites of winds and collimated jets are probed by Balmer and forbidden metallic emission lines.
Study of metallic emission lines, which all are velocity modulated due to
the rotation of the star, shows that
the narrow line components are produced in the post-shock region, while the broad components
originate in the more extended, pre-shock material in the accretion column.  Analysis of the time-dependent detailed profiles of emission lines associated with the accretion columns provide
insight into the details of the mass transfer due to magnetospheric accretion \citep{Sicilia-Aguilar2015}

\subsection{Disk Winds, Jets, and Outflows}

Magnetohydrodynamic (MHD) and photoevaporative winds are thought to play an important role in the evolution and dispersal of planet-forming disks \citep{Fang2018}.
These two wind mechanisms, along with possible MRI-driven turbulence, produce observational signatures due to structures in the outflow and the location of the wind launching radius. The dynamical feedback of
magneto-hydrodynamic disk winds on the planet formation zone induces
fast radial accretion and modification of planet migration \citep{Frank2014}.

The existence and character of magnetothermal winds are also uncertain. Theoretical
studies predict that winds capable of driving disk accretion at the observed stellar accretion rates will be massive, with mass loss rates comparable to disk accretion rates. It has
been suggested that the low velocity component of the [OI] 6300{\AA} line emission from T Tauri
stars provides evidence for magnetothermal winds \citep{Simon2016}. However, the decomposition
of a complex [OI] 6300{\AA} profile into multiple components potentially introduces
uncertainty in the interpretation. More detailed studies of this and other diagnostics, combined
with quantitative theoretical predictions of observable wind signatures can potentially
verify the existence and angular momentum transport properties of magnetothermal winds.

MHD winds produce spiral structure in the outflows which present possible observational signatures that could 
detected similarly to spectroastrometry, targeting narrow forbidden O lines using multiple observations with different slit rotations.
Constraining the wind launching radius by, e.g., identifying
the Keplerian component to the line profile would also help distinguish between photoevaporation and magnetocentrifugal winds. 

3-D modeling of non-ideal MHD disks \citep{Suriano2019} reveals that disk and wind turbulence is suppressed by ambipolar diffusion with the result that laminar flow
conditions likely dominate in the planet-forming region of 1--30 AU\citep{Banzatti2019}. However, neither 
mechanism (MRI or magnetothermal winds) has a verified observational signature although
sisk turbulence possibly driven by the MRI has been
detected both within 1 AU and beyond 40 AU. 
Within radii of 0.3 AU, high resolution
spectroscopy of CO overtone emission has uncovered evidence for non-thermal velocities
comparable to the sound speed in a few disk atmospheres consistent with the non-thermal
motions expected for MRI-driven turbulence. 
Measuring the turbulence on the surface at {\bf $\sim$1--5 AU} could help tell the difference between angular momentum transport by MRI and MHD winds \citep{Najita2018Disks}.

Emission line profiles result from the contributions of large scale bulk motion, e.g. from Keplerian orbits or 
outflows, thermal motion, and turbulence. 
Within the disk, the non-Keplerian contributions to the
line broadening are, following \citet{Teague2016}, 
\begin{equation}
\begin{split}
(\Delta v)^2 & = \frac{2 k T_{kin}}{\mu m_{H}}  +  \delta v^2_{turb}.
\end{split}
\end{equation}

The use of relatively heavy molecules reduces the thermal component of the motion, for example millimeter-wave studies of CO \citet{Flaherty2018} find 
limits on the turbulent motion in the disk of TW Hya of $\sim 0.08 c_{s}$, or 30 m s$^{-1}$ at 140 AU.  
Elements heavier than CO, and thus would be suited for 
turbulence studies, that give rise to forbidden lines found in T Tauri disk optical spectra include Ca and Fe.

\subsection{Jet Interactions with the Environment: Shocks} 

\begin{figure}[ht]
\begin{center}
\includegraphics[width=0.6\textwidth]{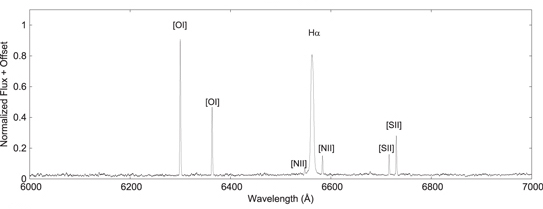}
\caption{
From \citet{Riaz2015}. 
A portion of the R$\sim$40,000 VLT/UVES spectra for HH 1158 with the prominent accretion- and outflow-associated emission lines marked. 
\label{fig:hh1158}}
\end{center}
\end{figure}

Warm outflow gas  ($10^3$--$10^4$ K) is heated by
internal shocks and moves at 10--100 km s$^{-1}$ with acceleration seen within the {\bf first 30 AU}
\citep{Schneider2014,Djupvik2016}. The resulting jet--ISM shocked interaction regions,
or Herbig--Haro objects, span
hundreds of AU. 
\'Echelle observations of shock tracers revealed component FWHMs of 15--20 km
s$^{-1}$. Detailed kinematic studies in optical forbidden lines such as
[OI] 6300{\AA}, [OI] 6363{\AA},
[NII]6584{\AA}, [SII] 6717{\AA}, and
[SII] 6731{\AA}
require resolutions better than 10 km s$^{-1}$ 
\citep{Davis2001}.

\section{Observations in the new era of ELTs}

While exoplanets have been discovered at distances ranging from within 0.01 AU to 100s of AU from their parent stars, the studies of circumstellar disks to date have only provided detailed views 
exterior to the planet forming region. New frontiers in the planet formation science will be
enabled by the high spectral and spatial resolution studies enabled by Extremely Large Telescopes
(ELTs) such as the Thirty Meter Telescope (TMT) and the Giant Magellan Telescope (GMT). 
The upcoming generation of ELTs will bridge this gap and allow us to investigate the initial conditions, locations, and timescales of planet formation in a way not previously possible. 

Optical spectroscopy with ELTS equipped with very high resolution spectrographs and ELTs of young stars and their disks benefits from the order of magnitude increase in sensitivity over existing
8 to 10 meter class facilities.  
Observations in the optical bands are constrained by natural seeing, with even Ground Layer Adaptive Optics giving no better than $\sim$0.25 to 0.40 arcseconds FWHM  in the 400--1000 nm range. At the 140 pc distance to the Taurus SFR, this gives spatial resolutions of 35--60 AU.  Spectra obtained centered on  
the young star will include lines from accretion, jets, and disk winds, providing a wealth of information
about the immediate circumstellar environment and the planet forming environment.

The study of emission lines giving state of the art
accretion and wind diagnostics drive spectral resolution requirements. These lines include both
broad- and narrow-line features requiring R $> 50,000$ to study line
profile variability and radial velocities 
\citep{Sicilia-Aguilar2015}. Key diagnostics of these processes are accessible in the optical bands with high resolution spectroscopy:
\begin{itemize}
\item Decomposition of emission line profiles showing structure 
within the accretion column.
\item Detection of companions from radial velocity signatures.
\item Details of the jet/ISM interaction.
\item Details on mass loss, momentum transport.
\item Substructure in the forbidden line HVC.
\item Details on the narrow component of the forbidden line LVC, including turbulence information in the disk wind.
\end{itemize}

The present state of the art is exemplified by optical echelle spectrographs such as HIRES on the 10-m Keck telescope [300-1000
nm, R up to 85K] and UVES on the 8.2-meter VLT [300-1100 nm, up to R ~ 80K on blue and 110K on red].
On the right side of Figure \ref{fig:example_lines} is shown example Keck/HIRES spectra at R $\sim$48K probing the disk wind and outflow
launching environments. An instrument such as TMT/HIRES or GMT/G-CLEF would 
enable similar observations at a factor of nine increase in sensitivity, giving access to the finest details
and temporal variability of the gas flow in these regions.

\pagebreak

\bibliographystyle{apj}
\begingroup
    \setlength{\bibsep}{0pt}
    \bibliography{references}

\begin{thebibliography}{}
\expandafter\ifx\csname natexlab\endcsname\relax\def\natexlab#1{#1}\fi

\bibitem[{{Banzatti} {et~al.}(2019){Banzatti}, {Pascucci}, {Edwards}, {Fang},
  {Gorti}, \& {Flock}}]{Banzatti2019}
{Banzatti}, A., {Pascucci}, I., {Edwards}, S., {et~al.} 2019, \apj, 870, 76

\bibitem[{{Davis} {et~al.}(2001){Davis}, {Hodapp}, \& {Desroches}}]{Davis2001}
{Davis}, C.~J., {Hodapp}, K.~W., \& {Desroches}, L. 2001, \aap, 377, 285

\bibitem[{{Djupvik} {et~al.}(2016){Djupvik}, {Liimets}, {Zinnecker}, {Barzdis},
  {Rastorgueva-Foi}, \& {Petersen}}]{Djupvik2016}
{Djupvik}, A.~A., {Liimets}, T., {Zinnecker}, H., {et~al.} 2016, \aap, 587, A75

\bibitem[{{Fang} {et~al.}(2018){Fang}, {Pascucci}, {Edwards}, {Gorti},
  {Banzatti}, {Flock}, {Hartigan}, {Herczeg}, \& {Dupree}}]{Fang2018}
{Fang}, M., {Pascucci}, I., {Edwards}, S., {et~al.} 2018, \apj, 868, 28

\bibitem[{{Flaherty} {et~al.}(2018){Flaherty}, {Hughes}, {Teague}, {Simon},
  {Andrews}, \& {Wilner}}]{Flaherty2018}
{Flaherty}, K.~M., {Hughes}, A.~M., {Teague}, R., {et~al.} 2018, \apj, 856, 117

\bibitem[{{Frank} {et~al.}(2014){Frank}, {Ray}, {Cabrit}, {Hartigan}, {Arce},
  {Bacciotti}, {Bally}, {Benisty}, {Eisl{\"o}ffel}, {G{\"u}del}, {Lebedev},
  {Nisini}, \& {Raga}}]{Frank2014}
{Frank}, A., {Ray}, T.~P., {Cabrit}, S., {et~al.} 2014, in Protostars and
  Planets VI, ed. H.~{Beuther}, R.~S. {Klessen}, C.~P. {Dullemond}, \&
  T.~{Henning}, 451

\bibitem[{{Haworth} {et~al.}(2016){Haworth}, {Ilee}, {Forgan}, {Facchini},
  {Price}, {Boneberg}, {Booth}, {Clarke}, {Gonzalez}, {Hutchison}, {Kamp},
  {Laibe}, {Lyra}, {Meru}, {Mohanty}, {Pani{\'c}}, {Rice}, {Suzuki}, {Teague},
  {Walsh}, {Woitke}, \& {Community authors}}]{Haworth2016}
{Haworth}, T.~J., {Ilee}, J.~D., {Forgan}, D.~H., {et~al.} 2016, Publications
  of the Astronomical Society of Australia, 33, e053

\bibitem[{{Najita} \& {Bergin}(2018)}]{Najita2018Disks}
{Najita}, J.~R., \& {Bergin}, E.~A. 2018, \apj, 864, 168

\bibitem[{{Pascucci} {et~al.}(2011){Pascucci}, {Sterzik}, {Alexander},
  {Alencar}, {Gorti}, {Hollenbach}, {Owen}, {Ercolano}, \&
  {Edwards}}]{Pascucci2011}
{Pascucci}, I., {Sterzik}, M., {Alexander}, R.~D., {et~al.} 2011, \apj, 736, 13

\bibitem[{{Riaz} \& {Whelan}(2015)}]{Riaz2015}
{Riaz}, B., \& {Whelan}, E.~T. 2015, \apj, 815, L31

\bibitem[{{Schneider} {et~al.}(2014){Schneider}, {Eisl{\"o}ffel}, {G{\"u}del},
  {G{\"u}nther}, {Herczeg}, {Robrade}, \& {Schmitt}}]{Schneider2014}
{Schneider}, P.~C., {Eisl{\"o}ffel}, J., {G{\"u}del}, M., {et~al.} 2014, in
  European Physical Journal Web of Conferences, Vol.~64, 08007

\bibitem[{{Sicilia-Aguilar} {et~al.}(2015){Sicilia-Aguilar}, {Fang},
  {Roccatagliata}, {Collier Cameron}, {K{\'o}sp{\'a}l}, {Henning},
  {{\'A}brah{\'a}m}, \& {Sipos}}]{Sicilia-Aguilar2015}
{Sicilia-Aguilar}, A., {Fang}, M., {Roccatagliata}, V., {et~al.} 2015, \aap,
  580, A82

\bibitem[{{Sicilia-Aguilar} {et~al.}(2016){Sicilia-Aguilar}, {Banzatti},
  {Carmona}, {Stolker}, {Kama}, {Mendigut{\'\i}a}, {Garufi}, {Flaherty}, {van
  der Marel}, \& {Greaves}}]{Sicilia-Aguilar2016}
{Sicilia-Aguilar}, A., {Banzatti}, A., {Carmona}, A., {et~al.} 2016,
  Publications of the Astronomical Society of Australia, 33, e059

\bibitem[{{Simon} {et~al.}(2016){Simon}, {Pascucci}, {Edwards}, {Feng},
  {Gorti}, {Hollenbach}, {Rigliaco}, \& {Keane}}]{Simon2016}
{Simon}, M.~N., {Pascucci}, I., {Edwards}, S., {et~al.} 2016, \apj, 831, 169

\bibitem[{{Suriano} {et~al.}(2019){Suriano}, {Li}, {Krasnopolsky}, {Suzuki}, \&
  {Shang}}]{Suriano2019}
{Suriano}, S.~S., {Li}, Z.-Y., {Krasnopolsky}, R., {Suzuki}, T.~K., \& {Shang},
  H. 2019, \mnras, 484, 107

\bibitem[{{Teague} {et~al.}(2016){Teague}, {Guilloteau}, {Semenov}, {Henning},
  {Dutrey}, {Pi{\'e}tu}, {Birnstiel}, {Chapillon}, {Hollenbach}, \&
  {Gorti}}]{Teague2016}
{Teague}, R., {Guilloteau}, S., {Semenov}, D., {et~al.} 2016, \aap, 592, A49

\end{thebibliography}
\endgroup

\end{document}